\documentclass[prd,twocolumn,floatfix,preprintnumbers,letterpaper]{revtex4}
\usepackage{graphics}
\usepackage{epsfig}
\usepackage{amsmath}

\def\4he{$^4$He}

\begin{document}

\title{The Uncertainty in Newton's Constant and Precision
Predictions of the Primordial Helium Abundance}
\author{Robert J. Scherrer}
\affiliation{Department of Physics and Astronomy, Vanderbilt University,
Nashville, TN  ~~37235}

\begin{abstract}
The current uncertainty in Newton's constant, $G_N$, is of the order of 0.15\%.
For values of the baryon to photon ratio consistent with both cosmic
microwave background observations and the primordial deuterium abundance, this
uncertainty in $G_N$
corresponds to an uncertainty in the primordial $^4$He mass fraction,
$Y_P$, of $\pm 1.3 \times 10^{-4}$.  This uncertainty in $Y_P$ is comparable to the
effect from the current uncertainty in the neutron lifetime, $\tau_n$,
which is often treated as the dominant uncertainty in calculations of $Y_P$.
Recent measurements of $G_N$ seem to be converging within a smaller range;
a reduction in the estimated error on $G_N$ by a factor of 10 would
essentially eliminate it as a source of uncertainty the calculation of the
primordial $^4$He abundance.

\end{abstract}

\maketitle

Big Bang nucleosynthesis (BBN) represents one of the key successes of modern
cosmology \cite{Rev1,Rev2}.  In recent years, the BBN production of
deuterium has emerged as
the most useful constraint on the baryon density in the universe, both because of
observations of deuterium in (presumably unprocessed) high-redshift QSO absorption line
systems
(see Ref. \cite{Kirkman}, and references therein), and the fact that the predicted BBN yields of deuterium are highly sensitive
to the baryon density.    These arguments give a baryon density parameter
$\Omega_b h^2$, of \cite{Kirkman}
\begin{equation}
\Omega_b h^2 = 0.0194 - 0.0234,
\end{equation}
which is in excellent agreement with the baryon density derived by the WMAP team
from recent cosmic microwave background observations 
\cite{Spergel}:
\begin{equation}
\Omega_b h^2 = 0.022 - 0.024.
\end{equation}
Given these limits on the baryon density, BBN predicts the primordial abundances
of $^4$He and $^7$Li.
Because the $^4$He abundance is particularly sensitive to new physics beyond
the standard model, a comparison between the predicted and observed abundances of $^4$He
can be used to constrain, for example, neutrino degeneracy or extra
relativistic degrees of freedom (see, e.g., Refs. \cite{Kneller,Barger}).

For this reason, it is useful to obtain the most accurate possible theoretical predictions
for the primordial $^4$He mass fraction, $Y_P$.
The primordial production of \4he is controlled by the competition between
the rates for the processes which govern the interconversion of neutrons
and protons,
\begin{eqnarray}
n+\nu_e &\leftrightarrow& p+e^-,\nonumber \\
n+e^+&\leftrightarrow& p+\bar \nu_e,\nonumber \\
n&\leftrightarrow& p+e^-+\bar\nu_e,
\end{eqnarray}
and the expansion rate of the Universe, given by
\begin{equation}
\label{H}
\frac{\dot R}{R} = \left(\frac{8}{3} \pi G_N \rho\right)^{1/2}.
\end{equation}
In BBN calculations, the weak interaction rates are scaled off of the inverse of the neutron lifetime, $\tau_n$.
When these
rates are faster than the expansion rate, the neutron-to-proton
ratio ($n/p$) tracks its equilibrium value.  As the Universe expands and
cools, the expansion rate comes to dominate and $n/p$ essentially freezes out. 
Nearly all the neutrons
which survive this freeze-out are bound into \4he when deuterium
becomes stable against
photodisintegration.
Following the initial
calculations of Wagoner, Fowler, and Hoyle \cite{Wagoner}, numerous groups examined
higher-order corrections to the $^4$He production.
The first such systematic attempt was undertaken by Dicus et al. \cite{Dicus},
who examined the effects of Coulomb and radiative corrections to the weak
rates,
finite-temperature QED effects, and incomplete neutrino coupling.
Later investigations included more detailed examination of 
Coulomb and radiative
corrections to the weak rates \cite{Donoghue1,Donoghue2,Sawyer,Chapman},
finite-temperature QED effects \cite{Heckler},
and incomplete neutrino decoupling \cite{Dodelson},
as well as an examination of the effects of finite nuclear mass \cite{Gyuk}.
These effects were systematized by
Lopez and Turner \cite{Lopez} (see also Ref. \cite{Esposito}), who argued that all
theoretical corrections larger than the effect of the uncertainty in the neutron
lifetime had been accounted for, yielding a total theoretical uncertainty
$\Delta Y_P < 0.0002$.  Assuming an uncertainty of $\pm 2$ sec in the neutron lifetime,
the corresponding experimental uncertainty in $Y_P$ is $\Delta Y_P = \pm 0.0004$.  Similar results
were obtained in Ref. \cite{Esposito}.

Two relevant changes have occurred since the publication of Refs. \cite{Lopez,Esposito}.  First,
the
estimated uncertainty in the neutron lifetime has decreased, with the current value
being \cite{PDG}
\begin{equation}
\label{tau}
\tau_n = 885.7 \pm 0.8 ~{\rm sec}.
\end{equation}
Second, the estimated uncertainty in
the value of $G_N$ has {\it increased}.  The value currently recommended by CODATA
(Committee on Data for Science and Technology) is \cite{Mohr}
\begin{equation}
\label{G}
G_N = 6.673 \pm 0.010 \times 10^{-8} {\rm cm}^3 {\rm gm}^{-1} {\rm sec}^{-2}.  
\end{equation}
This represents a factor of twelve increase over the
previous recommended uncertainty \cite{Cohen}, and is primarily due to an
anomalously
high value for $G$ determined by Michaelis, Haars, and Augustin \cite{Michaelis}.
(See Table 1).

\begin{table}

\begin{tabular}{|l|l|}
\hline \hline
$G_N \cdot 10^{8}$ [cm$^3$ g$^{-1}$ sec$^{-2}$] & Reference\\
\hline \hline
6.67407(22) & Schlamminger et al. (2002) \cite{Schlamminger}\\
6.67559(27) & Quinn, et al. (2001) \cite{Quinn}\\
6.674215(92) & Gundlach and Merkowitz (2000) \cite{Gundlach}\\
6.6699(7) & Luo et al. (1999) \cite{Luo}\\
6.6742(7) & Fitzgerald and Armstrong (1999) \cite{Fitzgerald}\\
6.6873(94) & Schwarz et al. (1999) \cite{Schwarz}\\
6.6749(14) & Nolting, et al. (1999) \cite{Nolting}\\
6.6735(29) & Kleinevoss et al. (1999) \cite{Kleinevoss}\\
6.673(10) & CODATA (1998) \cite{Mohr}\\
6.67259(85) & CODATA (1986) \cite{Cohen}\\
\hline \hline
\end{tabular}
\caption{Recent experimental values of Newton's constant.  Digits
in parentheses are the 1-$\sigma$ uncertainty in the last digits of the given
value.}
\end{table}

It is easy to calculate the effect of both of these uncertainties on the primordial
$^4$He abundance.  For the range of values of $\Omega_b h^2$ given in equations
(1) and (2), we find, numerically,
\begin{eqnarray}
\Delta Y_P &=& 0.088 (\Delta G_N/G_N),\\
\Delta Y_P &=& 0.18 (\Delta \tau_n/\tau_n).
\end{eqnarray}
The coefficient in equation (8) is twice that in equation (7).  This factor of 2
comes from the fact that
the expansion rate (equation \ref{H}) scales as $G_N^{1/2}$, while
the weak interaction rates scale as $\tau_n^{-1}$, and the abundance of $^4$He
is essentially unchanged if the ratio of the weak interaction rates to the
expansion rate is held constant.

Then the current uncertainties in $G_N$ and $\tau_n$ given in equations
(\ref{tau}) and (\ref{G}) yield, for the $1-\sigma$ uncertainties in $Y_P$,
\begin{equation}
\Delta Y_P = \pm 1.6 \times 10^{-4},
\end{equation}
from the uncertainty in $\tau_n$, and
\begin{equation}
\Delta Y_P = \pm 1.3 \times 10^{-4},
\end{equation}
from the uncertainty in $G_N$.

These two uncertainties are roughly comparable.  This is significant because
the uncertainty in $\tau_n$ is often taken to be the dominant uncertainty
in calculations of $Y_P$.  Of course, both of these effects are
exceedingly small, and well below the dispersion in the estimates of
the primordial $^4$He abundance from observations
of low-metallicity systems \cite{Rev2}.
(The effect
of the uncertainty in $G_N$ is comparable to the corrections to $Y_P$ due
to QED plasma effects and residual neutrino heating \cite{Lopez}).
A further source of uncertainty in theoretical calculations of $Y_P$ is the
uncertainty in the nuclear reaction rates.  A recent exhaustive study of this
effect has been undertaken by Cyburt \cite{Cyburt}, who concluded that the
effect on $Y_P$ of the uncertainties in the nuclear reaction rates
is currently subdominant.  (See also earlier work in Refs.
\cite{Krauss1,Krauss2}).

The uncertainty in $G_N$ has consequences in other astrophysical
settings.  Lopes and Silk \cite{Lopes} investigated the effect on
the sound speed in the sun.  They argued that helioseismology (in combination
with improved solar neutrino measurements) might eventually provide an
independent constraint on $G_N$, although this claim has been disputed by
Ricci and Villante \cite{Ricci}.

In principle, the uncertainty in $G_N$ also affects
CMB measurements.  The change in the observed CMB fluctuation spectrum due to a
fixed change in $G_N$ was investigated by Zahn and Zaldarriaga \cite{Zahn}.
Even under the most optimistic conditions for future observations,
the smallest change in $G_N$ which is, in principle, detectable in CMB measurements
is $\Delta G_N /G_N \sim 0.006$, well above the current CODATA uncertainty.

It is likely, of course, that current and future measurements will lead to a reduction in
the uncertainty in $G_N$.
A set of the most recent measurements of $G_N$
is displayed in Table 1.  (For a survey of measurements of $G_N$ over a longer
timeline, see Ref. \cite{Gillies}).  The three most recent measurements all yield a value
of $G_N$ within a very narrow range.
Reduction in the uncertainty in $G_N$ by, for example, a factor of 10 (e.g., back to the
1986 CODATA level of uncertainty) would essentially eliminate any significant effect on BBN
calculations.

\acknowledgments

I thank M. Kaplinghat and G. Greene for helpful discussions.

\newcommand\AJ[3]{~Astron. J.{\bf ~#1}, #2~(#3)}
\newcommand\APJ[3]{~Astrophys. J.{\bf ~#1}, #2~ (#3)}
\newcommand\apjl[3]{~Astrophys. J. Lett. {\bf ~#1}, L#2~(#3)}
\newcommand\ass[3]{~Astrophys. Space Sci.{\bf ~#1}, #2~(#3)}
\newcommand\cqg[3]{~Class. Quant. Grav.{\bf ~#1}, #2~(#3)}
\newcommand\mnras[3]{~Mon. Not. R. Astron. Soc.{\bf ~#1}, #2~(#3)}
\newcommand\mpla[3]{~Mod. Phys. Lett. A{\bf ~#1}, #2~(#3)}
\newcommand\npb[3]{~Nucl. Phys. B{\bf ~#1}, #2~(#3)}
\newcommand\plb[3]{~Phys. Lett. B{\bf ~#1}, #2~(#3)}
\newcommand\pr[3]{~Phys. Rev.{\bf ~#1}, #2~(#3)}
\newcommand\PRL[3]{~Phys. Rev. Lett.{\bf ~#1}, #2~(#3)}
\newcommand\PRD[3]{~Phys. Rev. D{\bf ~#1}, #2~(#3)}
\newcommand\prog[3]{~Prog. Theor. Phys.{\bf ~#1}, #2~(#3)}
\newcommand\RMP[3]{~Rev. Mod. Phys.{\bf ~#1}, #2~(#3)}


\begin{thebibliography}{99}

\bibitem{Rev1}
D.N. Schramm and M.S. Turner, Rev. Mod. Phys.
{\bf 70}, 303 (1998).

\bibitem{Rev2}
K.A. Olive, G. Steigman, and T.P. Walker,
Phys. Rep. {\bf 333}, 389 (2000).

\bibitem{Kirkman}
D. Kirkman, D. Tytler, N. Suzuki, J.M. O'Meara, and D. Lubin,
astro-ph/0302006.

\bibitem{Spergel}
D. Spergel, et al., Ap.J. Suppl. {\bf 148}, 175 (2003).

\bibitem{Kneller}
J.P. Kneller, R.J. Scherrer, G. Steigman, and T.P. Walker,
\prd {\bf 64}, 123506 (2001).

\bibitem{Barger}
V. Barger, J.P. Kneller, H.-S. Lee, D. Marfatia, and G. Steigman,
Phys. Lett. B {\bf 566}, 8 (2003).

\bibitem{Wagoner}
R.V. Wagoner, W.A. Fowler, and F. Hoyle, Ap.J. {\bf 148}, 3 (1967).

\bibitem{Dicus}
D.A. Dicus, et al, \prd {\bf 26}, 2694 (1982).

\bibitem{Donoghue1}
J.F. Donoghue and B.R. Holstein, \prd {\bf 28}, 340 (1983).

\bibitem{Donoghue2}
J.F. Donoghue and B.R. Holstein, \prd {\bf 29}, 3004 (1984).

\bibitem{Sawyer}
R.F. Sawyer, \prd {\bf 53}, 4232 (1996).

\bibitem{Chapman}
I.A. Chapman, \prd {\bf 55}, 6287 (1997).

\bibitem{Heckler}
A.J. Heckler, \prd {\bf 49}, 611 (1994).


\bibitem{Dodelson}
S. Dodelson and M.S. Turner, \prd {\bf 46}, 3372 (1992).

\bibitem{Gyuk}
R.E. Lopez,  M.S. Turner, and G. Gyuk, \prd {\bf 56}, 3191 (1997).

\bibitem{Lopez}
R.E. Lopez and M.S. Turner, \prd {\bf 59}, 103502 (1999).

\bibitem{Esposito}
S. Esposito, G. Mangano, G. Miele, and O. Pisanti,
Nucl. Phys. {\bf B540}, 3 (1999).

\bibitem{PDG}
K. Hagiwara, et al., \prd {\bf 66}, 010001 (2002).

\bibitem{Mohr}
P.J. Mohr and B.N. Taylor, Rev. Mod. Phys. {\bf 72}, 351 (2000).

\bibitem{Cohen}
E.R. Cohen and B.N. Taylor, Rev. Mod. Phys. {\bf 59}, 1121 (1987).

\bibitem{Michaelis}
W. Michaelis, H. Haars, and R. Augustin, Metrologia {\bf 32}, 267 (1996).

\bibitem{Cyburt}
R.H. Cyburt, astro-ph/0401091.

\bibitem{Krauss1}
L.M. Krauss and P. Romanelli, Ap.J. {\bf 358}, 47 (1990).

\bibitem{Krauss2}
L.M. Krauss and P. Kernan, Phys. Lett. {\bf B347}, 347 (1995).

\bibitem{Lopes}
I.P. Lopes and J. Silk, astro-ph/0112310.

\bibitem{Ricci}
B. Ricci and F.L. Villante, Phys. Lett. B {\bf 549}, 20 (2002).

\bibitem{Zahn}
O. Zahn and M. Zaldarriaga, \prd {\bf 67}, 063002 (2003).

\bibitem{Gillies}
G.T. Gillies, Meas. Sci. Technol. {\bf 10}, 421 (1999).

\bibitem{Schlamminger}
St. Schlamminger, E. Holzschuh, and W. K\"udig, \prl {\bf 89}, 161102 (2002).

\bibitem{Quinn}
T.J. Quinn, C.C. Speake, S.J. Richman, R.S. Davis, and A. Picard,
\prl {\bf 87}, 111101 (2001).

\bibitem{Gundlach}
J.H. Gundlach and S.M. Merkowitz, \prl {\bf 85}, 2869 (2000).

\bibitem{Luo}
Luo, J., Hu, Z.K., Fu X.H., Fan S.H., and Tang, M.X., \prd {\bf 59}, 042001 (1999).

\bibitem{Fitzgerald}
M.P. Fitzgerald and T.R. Armstrong, Meas. Sci. Technol {\bf 10}, 439 (1999).

\bibitem{Schwarz}
J.P. Schwarz, D.S. Robertson, T.M. Niebauer, and J.E. Faller, Meas.
Sci. Technol. {\bf 10}, 478 (1999).

\bibitem{Nolting}
F. Nolting, J. Schurr, S. Schlamminger, and W. K\"undig, Meas. Sci.
Technol. {\bf 10}, 487 (1999).

\bibitem{Kleinevoss}
U. Kleinevoss, H. Meyer, A. Schumacher, and S. Hartmann, Meas. Sci. Technol.
{\bf 10},  492 (1999).


\end{thebibliography}
\end{document}